\documentclass[review,preprint,12pt]{elsarticle}

\usepackage{amsmath,amssymb}
\usepackage{graphicx,lineno}
\usepackage{geometry,natbib}
\biboptions{numbers,sort&compress}
\usepackage[breaklinks,colorlinks=true,citecolor={blue}, linkcolor= {blue}, urlcolor ={blue}]{hyperref}

\begin{document}

\begin{frontmatter}

\title{{\bf Significant Mobility Enhancement in Coupled AlGaN/GaN Quantum Wells considering Inter-Well Distance and Asymmetric Widths}}

\author[l1,l2]{Le Tri Dat}
\address[l1]{Engineering Research Group, Dong Nai Technology University, Bien Hoa City, Vietnam}
\address[l2]{Faculty of Engineering, Dong Nai Technology University, Bien Hoa City, Vietnam.} \ead{letridat@dntu.edu.vn} 

\author[addr1]{Tran Trong Tai}
\author[addr1]{Truong Van Tuan}
\address[addr1]{University of Tran Dai Nghia, 189-Nguyen Oanh Street, Go Vap District, Ho Chi Minh City, Vietnam}

\author[addr2,addr3]{Vo Van Tai}
\address[addr2]{Laboratory of Applied Physics, Science and Technology Advanced Institute, Van Lang University, Ho Chi Minh City, Vietnam}
\address[addr3]{Faculty of Applied Technology, School of Technology, Van Lang University, Ho Chi Minh City, Vietnam}

\author[addr2,addr3]{Nguyen Duy Vy\corref{cor1}}
\cortext[cor1]{Corresponding author email: nguyenduyvy@vlu.edu.vn}

\begin{abstract}
We demonstrate that coupled AlGaN/GaN quantum wells with asymmetric widths ($L_1-L_2<30 $ {\AA} achieve up to 4.5× higher mobility than single wells at optimal separation ($d$ = 100 {\AA}). Crucially, mobility surpasses single wells when $d>40$ {\AA} reversing the trend at smaller distances. This enhancement stems from double-layer screening that suppresses remote/background impurities and dislocations, while LO phonon scattering remains unaffected. For identical wells, coupled systems underperform single wells at $d<40$ {\AA} but exceed them beyond this threshold. Peak gains occur at cryogenic temperatures (77 K). Our results provide a robust theoretical framework to optimize mobility in AlGaN/GaN heterostructures, reducing experimental trial-and-error in quantum device engineering.
\end{abstract}


\end{frontmatter}

\section{Introduction}
Extensive theoretical and experimental research has delved into calculating and measuring mobility in infinite square quantum wells, particularly those fabricated from the wide‑bandgap heterostructure $\mathrm{AlGaN/GaN/AlGaN}$, as reported in Refs. \cite{Zakhleniuk1999,Maeda2000,Maeda2001,Anderson2001,Anderson2003,Chen2003,Meng2012,Gu2017,Tuan2021}.
 Researchers, driven by the quest to enhance mobility in infinite square quantum wells of $\mathrm{AlGaN/GaN/AlGaN}$, have persistently sought innovative solutions. To achieve the highest possible mobility, it is essential to compare the mobilities of finite and infinite quantum wells, as meticulously calculated by the authors \cite{Tuan2025,Tai2025}. This comparison sheds light on significant insights that can drive advancements in the field. Besides, a noteworthy approach has involved investigating the combination of various semiconductors, contributing valuable insights into both theoretical models and the measurement of transport properties \cite{Smith2003,Hosono2014,Vazifehshenas2015,Linh2020,Tuan2023,Tuan2024}. This ongoing exploration is crucial for advancing the performance of quantum well structures. The authors' research reveals that there has been a notable lack of calculations concerning the mobility of coupled infinite square quantum wells, especially in the context of the $\mathrm{AlGaN/GaN/AlGaN}$ infinite square quantum well. When two quantum wells are coupled, the second layer significantly influences the first layer's mobility through the screening function. In this study, we rigorously investigate strategies to enhance the mobility of the $\mathrm{AlGaN/GaN/AlGaN}$ infinite square quantum well system. By conducting a thorough comparison between the mobility of a single quantum well and that of the coupled system—based on the $\mathrm{AlGaN/GaN/AlGaN}$ material—we aim to uncover effective theoretical pathways for increasing mobility. Our findings could have a profound impact on the design and optimization of quantum well systems for advanced technological applications. 

\begin{figure}    \centering
\includegraphics[width=0.6\linewidth]{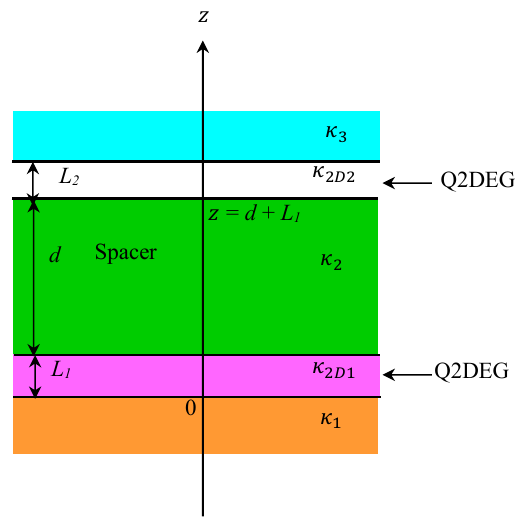}
\caption{A Q2DEG-Q2DEG double layer structure with different base dielectric layers}
    \label{fig:enter-label}
\end{figure}
 
We rigorously explore a range of scattering mechanisms that significantly impact mobility, including static potential scattering from remote ionized impurities (RI), homogeneous background charged impurities (BI), interface roughness (IR), and dislocation scattering, which can occur through either Coulomb interactions (DC) or strain fields (DS), and acoustic phonon scattering through deformation potential (acDP) and piezoelectric potential (acPE). Additionally, we focus on longitudinally polarized optical (LO) phonon scattering by employing an advanced iterative method to accurately calculate the inelastic region. Our analysis also delves into the crucial relationship between mobility and factors such as temperature, density, and well width in both identical and different quantum wells. Notably, we reveal that pairing two infinite square quantum wells composed of $\mathrm{AlGaN/GaN/AlGaN}$ leads to a significant enhancement in theoretical mobility. This comprehensive study underscores the potential for optimizing mobility in quantum well systems, paving the way for advancements in semiconductor technology.

\section{Theory}
We investigate a quasi-two-dimensional electron gas, referred to as Q2DEG, characterized by an effective mass $m^*$  and confined within an infinite square quantum well of width $L$ (where $0 < z < L$). At low temperatures and under conditions of low carrier density, it is essential to consider that the electrons predominantly occupy the lowest energy subband. The wavefunction in the $z$ direction can be described in the form  \cite{Gold1987,Tai2015,Tuan2021,Tuan2023b},
\begin{equation}
\Psi(z) = 
\begin{cases}
\sqrt{\dfrac{2}{L}} \sin\left( \dfrac{\pi z}{L} \right), & 0 \leq z \leq L \\
0, & z < 0 \ \text{or} \ z > L.
\end{cases}
\end{equation}
The electron mobility, as defined by the formal approximation of Boltzmann transport theory, is articulated in the following manner  \cite{Kawamura1992,Hamaguchi2017},
\begin{equation}
\mu = \dfrac{e}{m^*} \langle \tau_{\text{tot}}(E) \rangle,
\end{equation}
where,
\begin{equation}
\langle \tau_{\text{tot}}(E) \rangle =
\dfrac{ \int_0^{+\infty} E \dfrac{\partial f_0(E)}{\partial E} \tau(E) \, \text{d}E }
     { \int_0^{+\infty} E \dfrac{\partial f_0(E)}{\partial E} \, \text{d}E },
\end{equation}
with $\tau(E)$ signifying the electron relaxation time for elastic scattering, and $f_0(E)$ indicating the Fermi-Dirac distribution function. The total relaxation time is clearly defined by the following relation:
\begin{equation}
\dfrac{1}{\tau_{\text{tot}}(E)} = \dfrac{1}{\tau_{\text{def}}(E)} + \dfrac{1}{\phi(E)},
\end{equation}
where $\tau_{\text{def}}(E)$ signifies the electron relaxation time arising from defect scattering, while $\phi(E)$ represents the perturbation distribution function influenced by phonon scattering  \cite{Kamatagi2005,Begum2014}. This distinction highlights the critical roles that both defect and phonon interactions play in the behavior of electrons, ultimately affecting material properties.

In Boltzmann transport theory, the concept of relaxation time is crucial and is articulated as follows \cite{Gold1987,Kawamura1992,Hamaguchi2017,Tai2015,Tuan2023b}. This definition serves as a fundamental element in understanding the dynamics of particle transport.
\begin{equation}
\frac{1}{\tau(E)} = \frac{m^*}{\pi \hbar^3} \int_0^{\pi} d\theta \, (1 - \cos \theta) \, \frac{\langle |U(q)|^2 \rangle}{\varepsilon^2(q,T)},
\end{equation}
where,
\begin{align}
\varepsilon(q,T) =& 1 + \frac{2\pi e^2}{\kappa_L}
\frac{1}{q} F_C(q) \Pi(q,T),
\\
\Pi(q,T) =& \frac{\beta}{4} \int_0^{\infty} d\xi' \, \frac{\Pi^0(q)}{\cosh^2\left[\frac{\beta}{2}(\xi - \xi')\right]},
\\
\Pi^0(q,E_F) =& \Pi^0(q) = \frac{m^*}{\pi \hbar^3} \left[1 - \sqrt{1 - \left(\frac{2k_F}{q}\right)^2} \, \Theta(q - 2k_F)\right],
\end{align}

\begin{align}
F_C(q) &= \int_0^L \! dz \int_0^L \! dz' \, |\Psi(z)|^2 |\Psi(z')|^2 e^{-q|z - z'|} \\
&= \frac{1}{4\pi^2 + L^2 q^2} \left[ 3Lq + \frac{8\pi^2}{Lq} 
+ \frac{32\pi^4 (e^{-Lq} - 1)}{L^2 q^2 (4\pi^2 + L^2 q^2)} \right],
\end{align}
with $\kappa_L$ being the dielectric constant, $\Pi^0(q)$ being the zero-temperature polarization function in the random phase approximation (RPA), $\beta = (k_B T)^{-1}$, $f_0(E) = \left[\exp\left(\frac{E - \xi}{k_B T}\right)+1\right]^{-1}$, $\xi = k_B T \ln\left[-1 + \exp\left(\frac{E_F}{k_B T}\right)\right]$, $E_F = \frac{\hbar^2 k_F^2}{2 m^*}$, $k_F = \sqrt{2\pi N_s}$, and $\langle |U(q)|^2 \rangle$ the random potential depending on the scattering mechanism.

In the context of remote ionized impurity scattering (RI), refer to sources \cite{Gold1987,Tai2015,Tuan2023b,Tuan2025,Tai2025},
\begin{equation}
\langle |U_{\text{RI}}(q)|^2 \rangle =
N_{\text{RI}}\!\left( \frac{2\pi e^2}{\bar{\kappa}\, q} \right)^{\!2}
\bigl[F_{\text{RI}}(q,z_i)\bigr]^{2}.
\end{equation}

When considering scattering due to homogeneous background charged impurities (BI), it is essential to refer to source \cite{Tuan2023b} for a comprehensive understanding:
\begin{equation}
\langle |U_{\text{BI}}(q)|^2 \rangle =
\left( \frac{2\pi e^2}{\bar{\kappa} q} \right)^2
\int_{-\infty}^{+\infty} dz_i \, N_{\text{BI}} \left[ F_{\text{RI}}(q,z_i) \right]^2
= \left( \frac{2\pi e^2}{\bar{\kappa} q} \right)^2 I_B.
\end{equation}
When examining interface roughness scattering (SR) \cite{Tai2015,Tuan2023b,Tuan2025,Tai2025}, it is essential to consider its significant implications and unique characteristics:
\begin{equation}
\langle |U_{\text{SR}}(q)|^2 \rangle = \left[ F_{\text{SR}}(q) \right]^2 \langle |\Delta(q)|^2 \rangle,
\end{equation}
with $\langle |\Delta(q)|^2 \rangle$ being the Fourier transform of the Gaussian autocorrelation function,
\begin{equation}
\langle |\Delta(q)|^2 \rangle = \pi \Delta^2 \Lambda^2 e^{-q^2 \Lambda^2 / 4}.
\end{equation}

Regarding Coulomb (DC) potential dislocation scattering, the significance of this phenomenon is well documented in sources \cite{Kamatagi2005,Tuan2023b},
\begin{equation}
|U_{\text{DC}}(q)|^2 =
N_{\text{dis}} \left( \frac{2\pi e^2 \rho_L}{\bar{\kappa} q} \right)^2
\left[ F_{\text{DC}}(q) \right]^2.
\end{equation}
For the potential resulting from strain field scattering (DS) \cite{Joshi2003,Tuan2023b},
\begin{equation}
\langle |U_{\text{DS}}(q)|^2 \rangle =
N_{\text{dis}}\, a_c^2 b_e^2 \left( \frac{1 - 2\gamma}{1 - \gamma} \right)^2
\frac{1}{2q^2} \left[ F_{\text{DS}}(q) \right]^2.
\end{equation}
In the context of acoustic deformation potential phonon scattering (acDP) \cite{Tuan2025,Tai2025,Tuan2023b,Hamaguchi2017}:
\begin{equation}
\langle |U_{\text{acDP}}^{\text{BG}}(q)|^2 \rangle =
\frac{1}{\pi} \int_0^{+\infty} dq_z \, |C_{\text{acDP}}(Q)|^2 |I(q_z)|^2 \Delta E.
\end{equation}
The piezoelectric acoustic phonon scattering (acPE) plays a pivotal role that cannot be overlooked:
\begin{equation}
\langle |U_{\text{acPE}(l,t)}^{\text{BG}}(q)|^2 \rangle =
\frac{1}{\pi} \int_0^{+\infty} dq_z \, |C_{\text{acPE}(l,t)}(Q)|^2 |I(q_z)|^2 \Delta E,
\end{equation}
where
$|C_{\text{acPE}(l,t)}(Q)|^2$ corresponds to a wurtzite crystal structure.

Longitudinally polarized optical phonon scattering (LO) is a crucial phenomenon that significantly impacts our understanding \cite{Tuan2025,Tai2025}:
\begin{equation}
\phi(E_k) =
\left[ S_0(E_k) \right]^{-1}
\left[
1 + S_a(E_k) \phi(E_k + \hbar \omega_Q)
+ S_e(E_k) \phi(E_k - \hbar \omega_Q)
\right].
\end{equation}
In the scenario of a double layer formed by two infinite square quantum wells, separated by a distance $d$ as depicted in Fig. 1, the dielectric function $1/\varepsilon^2(q,T)$ is effectively transformed into $\left( \frac{W_{ij}(q,T)}{V_{ij}(q)} \right)^2$. This transformation showcases the important role of the double-layer screening potential, denoted as $W_{ij}(q,T)$, where the indices $i$ and $j$ take on the values of 1 and 2, respectively \cite{Hu1993,Smith2003,Scharf2012,Jishi2013,Tuan2024,Tai2025b}. This change underscores the importance of understanding quantum interactions within this unique framework.
\begin{equation}
W_{ij}(q,T) = \frac{
V_{ii}(q) + \left[ V_{ii}(q)V_{jj}(q) - V_{ij}^2(q) \right] \Pi_{jj}(q,T)
}{
\left[1 + V_{ii}(q)\Pi_{ii}(q,T)\right] \left[1 + V_{jj}(q)\Pi_{jj}(q,T)\right]
- V_{ij}^2(q) \Pi_{ii}(q,T) \Pi_{jj}(q,T)
},
\end{equation}
with \( V_{ij}(q) \) being the Coulomb interaction potential:
\begin{equation}
V_{ij}(q) = \frac{2\pi e^2}{q} f_{ij}(q),
\end{equation}
and \( \Pi_{ii}(q,T) \) is the polarization function at temperature \( T \) of the \( i \)-th layer.  
For the Q2DEG–Q2DEG double layer system, \( f_{ij}(q) \) has the form, $x = qd$, $y = qL_1$, $z = qL_2$ \cite{Scharf2012,Tuan2024}.

The intralayer Coulomb interaction potential of quantum well 1 has the form:
\begin{equation}
f_{11}(x,y,z) = \frac{1}{D(x,y,z)} \left[
\frac{f_{11}^{(1)}(x,y,z)}{\kappa_{\mathrm{2D}1} \, y^2 (y^2 + 4\pi^2)^2}
+ \frac{f_{11}^{(2)}(x,y,z)}{y^2 (y^2 + 4\pi^2)^2}
+ \frac{f_{11}^{(3)}(x,y,z)}{y^2 (y^2 + 4\pi^2)^2}
\right],
\end{equation}
\begin{equation}
\begin{aligned}
f_{11}^{(1)}(x,y,z) &= \kappa_1 \kappa_2 \big[
\kappa_{2D2} \cosh(x) (\kappa_3 \cosh(z) + \kappa_{2D2} \sinh(z)) \\
&\quad + \kappa_2 \sinh(x) (\kappa_{2D2} \cosh(z) + \kappa_3 \sinh(z))
\big] \\
&\quad \times \left[
64\pi^4 (1 - \cosh(y)) + y(y^2 + 4\pi^2)(3y^2 + 8\pi^2)\sinh(y)
\right],
\end{aligned}
\end{equation}

\begin{align}
f_{11}^{(2)}(x,y,z) &= \cosh(x) \big[
\kappa_2 \kappa_{2D2} (\kappa_1 + \kappa_3) \cosh(z) 
+ \kappa_2 (\kappa_1 \kappa_3 + \kappa_{2D2}^2) \sinh(z)
\big]
\nonumber\\
&\quad + \sinh(x) \big[
\kappa_{2D2} (\kappa_1 \kappa_3 + \kappa_2^2) \cosh(z) 
+ (\kappa_1 \kappa_{2D2}^2 + \kappa_2^2 \kappa_3) \sinh(z)
\big]
\nonumber\\
&\quad \times \left[
y(y^2 + 4\pi^2)(3y^2 + 8\pi^2) \cosh(y) - 32\pi^4 \sinh(y)
\right],
\\
f_{11}^{(3)}(x,y,z) &= \kappa_{2D1} \big[
\kappa_2 \cosh(x)(\kappa_{2D2} \cosh(z) + \kappa_3 \sinh(z))
\nonumber\\
&\quad + \kappa_{2D2} \sinh(x)(\kappa_3 \cosh(z) + \kappa_{2D2} \sinh(z))
\big] 
\nonumber\\
&\quad \times \left[
y(y^2 + 4\pi^2)(3y^2 + 8\pi^2) \sinh(y)
\right],
\\
D(x, y, z) &= \kappa_2 \cosh(x) \left\{
    \kappa_{\mathrm{2D}1} \cosh(y) \left[
        \kappa_{\mathrm{2D}2} (\kappa_1 + \kappa_3) \cosh(z)
        + (\kappa_1 \kappa_3 + \kappa_{\mathrm{2D}2}^2) \sinh(z)
    \right] \right. \notag \\
&\quad \left. + \sinh(y) \left[
        \kappa_{\mathrm{2D}2} (\kappa_1 \kappa_3 + \kappa_{\mathrm{2D}1}^2) \cosh(z)
        + (\kappa_3 \kappa_{\mathrm{2D}1}^2 + \kappa_1 \kappa_{\mathrm{2D}2}^2) \sinh(z)
    \right]
\right\} \notag \\
&\quad + \sinh(x) \left\{
    \kappa_{\mathrm{2D}1} \cosh(y) \left[
        \kappa_{\mathrm{2D}2} (\kappa_1 \kappa_3 + \kappa_2^2) \cosh(z)
        + (\kappa_1 \kappa_{\mathrm{2D}2}^2 + \kappa_2^2 \kappa_3) \sinh(z)
    \right] \right. \notag \\
&\quad \left. + \sinh(y) \left[
        \kappa_{\mathrm{2D}2} (\kappa_1 \kappa_2^2 + \kappa_3 \kappa_{\mathrm{2D}1}^2) \cosh(z)
        + (\kappa_1 \kappa_2^2 \kappa_3 + \kappa_{\mathrm{2D}1}^2 \kappa_{\mathrm{2D}2}^2) \sinh(z)
    \right]
\right\}.
\end{align}

The intralayer Coulomb interaction potential of quantum well 2 has the form:
\begin{align}
f_{22}(x, y, z) &= \frac{1}{D(x, y, z)} \left[
    \frac{f_{22}^{(1)}(x, y, z)}{\kappa_{\mathrm{2D}2} z^2 (z^2 + 4\pi^2)^2}
    + \frac{f_{22}^{(2)}(x, y, z)}{z^2 (z^2 + 4\pi^2)^2}
    + \frac{f_{22}^{(3)}(x, y, z)}{z^2 (z^2 + 4\pi^2)^2}
\right],
\\
f_{22}^{(1)}(x,y,z) &= \kappa_2 \kappa_3 \big[
\kappa_{2D1} \cosh(x)(\kappa_1 \cosh(y) + \kappa_{2D1} \sinh(y))
\nonumber \\
&\quad + \kappa_2 \sinh(x)(\kappa_{2D1} \cosh(y) + \kappa_1 \sinh(y))
\big]
\nonumber\\
&\quad \times \left[
64\pi^4(1 - \cosh(z)) + z(z^2 + 4\pi^2)(3z^2 + 8\pi^2)\sinh(z)
\right],
\\
f_{22}^{(2)}(x,y,z) &= \cosh(x) \big[
\kappa_2 \kappa_{2D1}(\kappa_3 + \kappa_1)\cosh(y) + \kappa_2(\kappa_1 \kappa_3 + \kappa_{2D1}^2)\sinh(y)
\big]
\nonumber\\
&\quad + \sinh(x) \big[
\kappa_{2D1}(\kappa_1 \kappa_3 + \kappa_2^2)\cosh(y) + (\kappa_3 \kappa_{2D1}^2 + \kappa_2^2 \kappa_1)\sinh(y)
\big] 
\nonumber\\
&\quad \times \left[
z(z^2 + 4\pi^2)(3z^2 + 8\pi^2)\cosh(z) - 32\pi^4 \sinh(z)
\right],
\\
f_{22}^{(3)}(x,y,z) &= \kappa_{2D2} \big[
\kappa_2 \cosh(x)(\kappa_{2D1} \cosh(y) + \kappa_1 \sinh(y)) 
\nonumber\\
&\quad + \kappa_{2D1} \sinh(x)(\kappa_1 \cosh(y) + \kappa_{2D1} \sinh(y))
\big]
\nonumber\\
&\quad \times \left[
z(z^2 + 4\pi^2)(3z^2 + 8\pi^2)\sinh(z)
\right].
\end{align}

The cross-layer Coulomb interaction potential is calculated as follows:
\begin{equation}
f_{12}(x,y,z) = \frac{f_{12}^{(1)}(x,y,z)}{[y(y^2 + 4\pi^2)][z(z^2 + 4\pi^2)] D_{QQ}(x,y,z)},
\end{equation}
with,
\begin{align}
f_{12}^{(1)}(x,y,z) =& \kappa_2 \cdot 32\pi^4 \left[ \kappa_1(\cosh y - 1) + \kappa_{2D1} \sinh y \right] 
\left[ \kappa_3(\cosh z - 1) + \kappa_{2D2} \sinh z \right] ,
\\
f_{21}(x,y,z) =& \frac{f_{21}^{(1)}(x,y,z)}{[y(y^2 + 4\pi^2)][z(z^2 + 4\pi^2)] D_{QQ}(x,y,z)},
\end{align}
with,
\begin{equation}
f_{21}^{(1)}(x,y,z) = \kappa_2 \cdot 32\pi^4 \left[ \kappa_1(\cosh y - 1) + \kappa_{2D1} \sinh y \right] 
\left[ \kappa_3(\cosh z - 1) + \kappa_{2D2} \sinh z \right].
\end{equation}

\section{Numerical results}
In this section, we calculate numerical results for the mobility of Q2DEG in an infinite square quantum well, würtize AlGaN/GaN/AlGaN, and its double layer with the following parameters \cite{Tai2025,Kawamura1992,Begum2014,Kamatagi2005,VanTan2019}: $m^* = m_z = 0.22\,m_e$, $\rho = 6.15\,\mathrm{g/cm^3}$, $u_l = 6.56 \times 10^5\,\mathrm{cm\cdot s^{-1}}$, $u_t = 2.68 \times 10^5\,\mathrm{cm\cdot s^{-1}}$, $D = 8.3\,\mathrm{eV}$, $\kappa_L = 8.9$, $h_{33} = 10.86 \times 10^7\,\mathrm{V/cm}$, $h_{31} = -3.91 \times 10^7\,\mathrm{V/cm}$, $h_{15} = -3.57 \times 10^7\,\mathrm{V/cm}$, $N_{B1} = N_{B2} = N_{B3} = 10^{14}\,\mathrm{cm^{-3}}$, $N_{RI} = N_s$, $z_i = -L$, $a_c = -8.0\,\mathrm{eV}$, $b_e = a_0 = 3.189\,\text{\AA}$, $c_0 = 5.185\,\text{\AA}$, $\gamma = 0.3$, $f = 1$, $\Delta = 2\,\text{\AA}$, $\Lambda = 15\,\text{\AA}$, $N_{\text{dis}} = 10^8\,\mathrm{cm^{-2}}$, $\delta = 95\,\mathrm{meV}$ and $\varepsilon_{L\infty} = 5.47$; $\kappa_{2D1} = \kappa_{2D2} = \kappa_L$, $\kappa_1 = \kappa_2 = \kappa_3 = 1$.

\begin{figure}\centering
\includegraphics[width=1\linewidth]{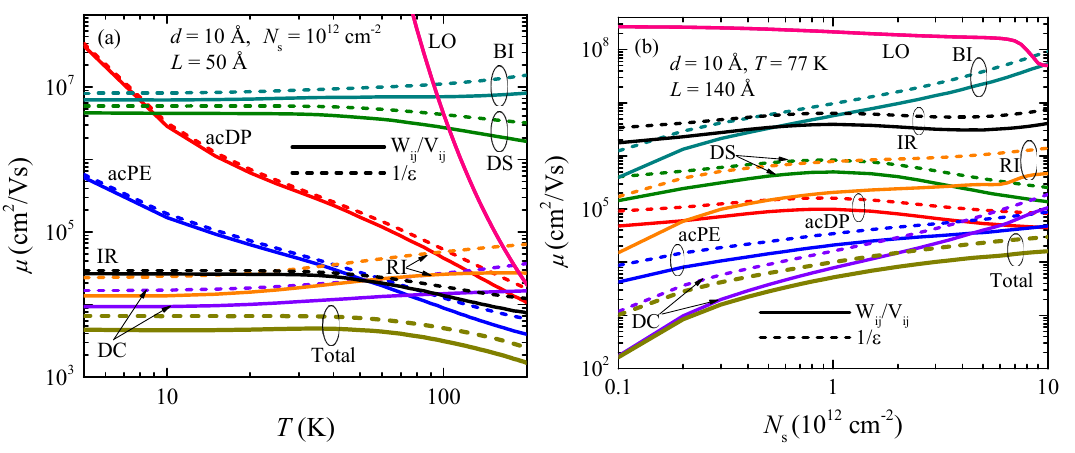}
\caption{Mobility of the coupled quantum well and the single quantum well corresponding to two screening functions $\left(\frac{W_{ij}(q,T)}{V_{ij}(q)}\right)^2$ and $1/\varepsilon^2(q,T)$ with different scattering mechanisms: (a) varying with temperature $T$; (b) varying with density $N_s$.
}
    \label{fig2}
\end{figure}

\begin{figure}    \centering
\includegraphics[width=1\linewidth]{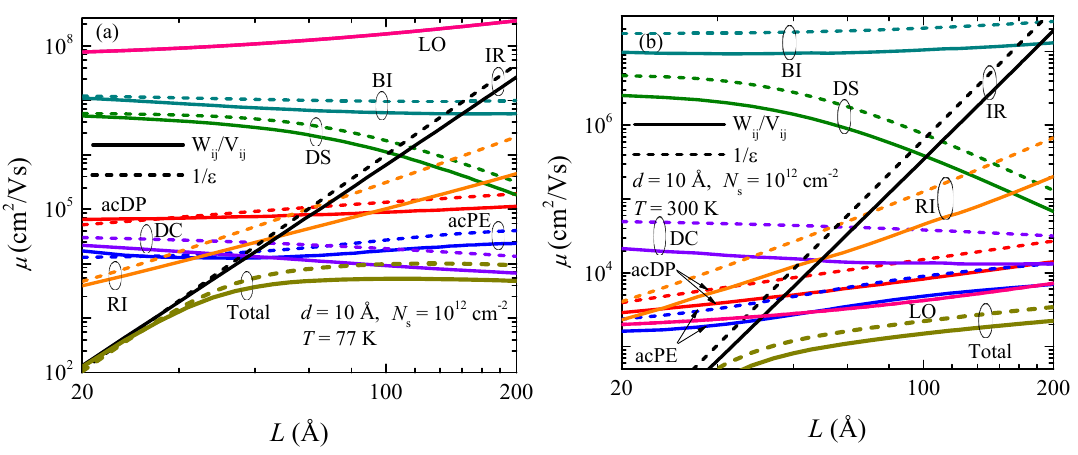}
    \caption{Mobility of single and double quantum wells with different scattering mechanisms depending on the well width L calculated according to the corresponding screening functions: a) $T$ = 77 K, b) $T$ = 300 K.}
    \label{fig3}
\end{figure}

\begin{figure}    \centering
\includegraphics[width=0.6\linewidth]{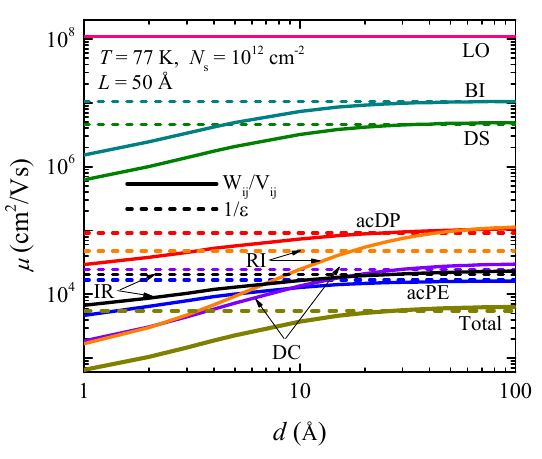}
    \caption{Mobility variation with distance $d$ between two quantum wells of scattering mechanisms with screening functions $\left(\frac{W_{ij}(q,T)}{V_{ij}(q)}\right)^2$ and $\frac{1}{\varepsilon^2(q,T)}$.}
    \label{fig4}
\end{figure}

\begin{figure}    \centering
\includegraphics[width=1\linewidth]{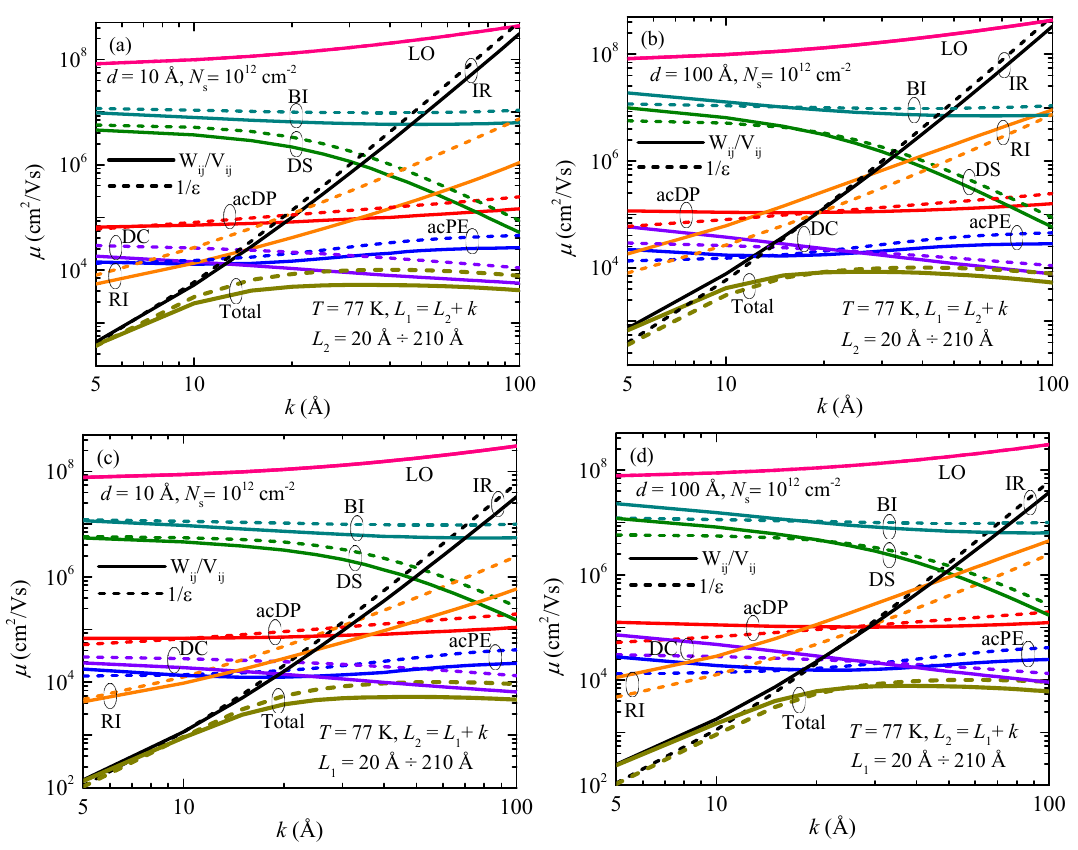}
\caption{Mobility of single quantum well with respect to the screening function \( \frac{1}{\varepsilon^2(q,T)} \), of double quantum well with respect to the screening function \( \left( \frac{W_{ij}(q,T)}{V_{ij}(q)} \right)^2 \) where \( L_1 = L_2 + k \) (\text{\AA}) with a) \( d = 10 \) \text{\AA}, b) \( d = 100 \) \text{\AA}; \( L_2 = L_1 + k \) (\text{\AA}) with c) \( d = 10 \) \text{\AA}, d) \( d = 100 \) \text{\AA}.}
    \label{fig5}
\end{figure}

\begin{figure}    \centering
\includegraphics[width=1\linewidth]{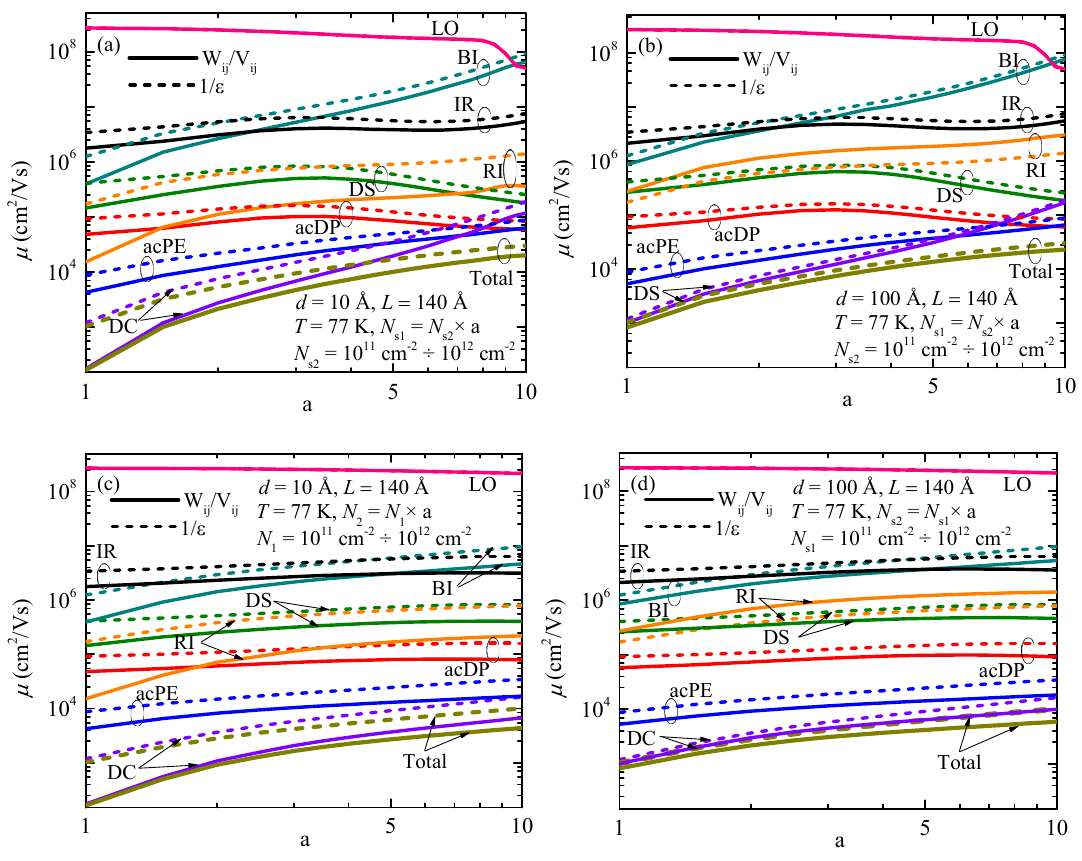}
\caption{Mobility of single quantum well with respect to the screening function \( \frac{1}{\varepsilon^2(q,T)} \), of double quantum well with respect to the screening function \( \left( \frac{W_{ij}(q,T)}{V_{ij}(q)} \right)^2 \) where \( N_1 = N_2 \times a \) with a) \( d = 10 \ \text{\AA} \), b) \( d = 100 \ \text{\AA} \); \( N_2 = N_1 \times a \) with c) \( d = 10 \ \text{\AA} \), d) \( d = 100 \ \text{\AA} \).}
    \label{fig6}
\end{figure}

In Figs. \ref{fig2} to \ref{fig4}, we analyze how mobility changes with variations in density and well width for two quantum wells in a double layer configuration. Then, in Figs. \ref{fig5} and \ref{fig6}, we explore how mobility varies with different density and well width for the same two quantum wells in a double layer and compare these results to those of a single quantum well.

Figure \ref{fig2} illustrates the mobility of a single quantum well in relation to the screening function $1/\varepsilon^2(q,T)$ and the mobility of a double quantum well concerning the screening function $\left(\frac{W_{ij}(q,T)}{V_{ij}(q)}\right)^2$. Specifically, Fig. \ref{fig2}(a) shows how mobility changes with temperature at $d = 10\,\text{\AA}$, $N_s = 10^{12}\,\text{cm}^{-2}$ and $L = 50\,\text{\AA}$. Figure~2b, on the other hand, depicts the change with density at $d = 10\,\text{\AA}$, $T = 77\,\text{K}$ and $L = 140\,\text{\AA}$. In polar semiconductors such as GaN, electron scattering by LO phonons is primarily described by the Fröhlich interaction, which arises from the oscillating polarization field generated by lattice ions. This polarization field oscillates at the LO phonon frequency. At such high frequencies, electrons cannot effectively respond or rearrange to screen the polarization field, rendering electron screening negligible. Consequently, the dynamic dielectric function approaches unity, indicating minimal screening by conduction electrons. As a result, the Fröhlich coupling strength remains essentially unscreened, independent of the electronic dielectric screening. Therefore, electron mobility limited by LO phonon scattering does not depend on screening effects in either single-layer or multilayer Q2DEG systems based on GaN \cite{Tuan2021,Tuan2025,Tai2025}. A significant difference is observed for BI, DS, RI, DC (acDP, acPE, IR) scattering at the examined temperatures (at high temperatures), particularly in Fig. \ref{fig2}(a). Meanwhile, in Fig. \ref{fig2}(b), the disparity in mobilities for all scattering mechanisms between the single and double quantum wells becomes even more pronounced at the specified density. Consequently, the total mobility of the single quantum well is always greater, nearly 1.6 times larger than that of the double quantum well. For instance, at $T = 200\,\text{K}$, $d = 10\,\text{\AA}$, $N_s = 10^{12}\,\text{cm}^{-2}$ and $L = 50\,\text{\AA}$, the ratio of mobilities is approximately $\mu^{\text{single}} / \mu^{\text{double}} \approx \frac{2526}{1570} \approx 1.6$.

In Fig. \ref{fig3}, we compellingly demonstrate the relationship between mobility and well width $L$ for a distance of $d = 10$~Å and a carrier density $N_s = 10^{12}$~cm$^{-2}$ at two key temperature values: $T = 77$~K and $T = 300$~K. At the lower temperature of $T = 77$~K, the mobility derived from the screening function $\frac{1}{\varepsilon^2(q,T)}$ stands out as significantly higher than that from the screening function $\left( \frac{W_{ij}(q,T)}{V_{ij}(q)} \right)^2$ for narrow well widths ($L > 30$~Å). This trend persists even for wider well widths ($L > 40$~Å), underscoring the critical role of well width in enhancing mobility. This phenomenon is further exemplified in Fig. \ref{fig5}. At the elevated temperature of $T = 300$~K, the mobility of scattering mechanisms, as indicated by the $\frac{1}{\varepsilon^2(q,T)}$ screening function, remains strikingly superior to the screening function $\left( \frac{W_{ij}(q,T)}{V_{ij}(q)} \right)^2$ across all examined well widths. Notably, under the conditions of $d = 10$~Å, $N_s = 10^{12}$~cm$^{-2}$, $T = 300$~K, we observe that $\mu^{\text{single}} \approx 1.5 \times \mu^{\text{double}}$. This evidence strongly highlights the significant impact of temperature and well width on mobility.

Figure \ref{fig4} clearly illustrates how the mobility of single and double quantum wells changes with the distance \( d \) between the two quantum wells, influenced by various scattering mechanisms. Notably, the mobility of the single quantum well remains consistent, showing no dependence on the distance \( d \). In the regime where \( d < 40~\text{\AA} \), the total mobility of the double quantum well is clearly inferior to that of the single quantum well. However, as \( d \) exceeds \( 40~\text{\AA} \), the total mobility of the double quantum well not only approaches but surpasses that of the single quantum well, showcasing a crucial shift. Focusing on RI scattering exclusively, it becomes evident that, for larger values of \( d \), the total mobility of the double quantum well significantly outperforms that of the single quantum well. This remarkable improvement can be attributed to the long-range nature of RI scattering, which diminishes in effect as \( d \) increases. To optimize the mobility of the system, strategically increasing the distance \(d\) between the two quantum wells while keeping the electron concentration and well width constant is key. Specifically, in the critical range of \(40~\text{\AA} < d < 100~\text{\AA}\), with a temperature of \(T = 77~\text{K}\), an electron concentration of \(N_s = 10^{12}~\text{cm}^{-2}\), and a well width \(L = 50~\text{\AA}\), we find the mobility values to be striking: \(5898~(\text{cm}^2/\text{Vs}) < \mu^{\text{double}} < 6355~(\text{cm}^2/\text{Vs}),\) and \(\mu^{\text{single}} = 5525~(\text{cm}^2/\text{Vs})\). This evidence underscores the significant advantages of optimizing quantum well structures for enhanced performance.

In Figs. \ref{fig5} and \ref{fig6}, we present an interesting method for increasing the mobility of a double quantum well by varying the density and width of the two individual quantum wells non-uniformly. Figure 5 illustrates the mobility of both single and double quantum wells corresponding to two screening functions: \( \frac{1}{\varepsilon^2(q,T)} \) and \( \left( \frac{W_{ij}(q,T)}{V_{ij}(q)} \right)^2 \). The single quantum well is modeled based on \( L_1 \), similar to the first quantum well in the double-layer system at a temperature of 77 K, with \( N_s = 10^{12} \ \text{cm}^{-2} \). We consider two values for the distance \( d \) between the two quantum wells, specifically \( d = 10 \ \text{\AA} \) and \( d = 100 \ \text{\AA} \). For Figures 5a and 5b, we have \( L_1 = L_2 + k \ (\text{\AA}) \), while Figures 5c and 5d show \( L_2 = L_1 + k \ (\text{\AA}) \). From these figures, we note that the LO scattering does not depend on the barrier function, resulting in no change in mobility. Additionally, the total mobility of the single quantum well is consistently greater than that of the double-layer quantum well within the examined well width region (refer to Figs. \ref{fig5}(a) and (c). However, when \( d \) is increased to \( 100 \ \text{\AA} \), the mobility of the single quantum well only maintains its superiority when \( k > 30 \ \text{\AA} \) for most scattering mechanisms, except for the RI scattering (see Figures 5b and 5d). Specifically, with the parameters \( k = 35 \ \text{\AA} \), \( L_1 = 115 \ \text{\AA} \), \( L_2 = 80 \ \text{\AA} \), \( d = 100 \ \text{\AA} \), \( N_s = 10^{12} \ \text{cm}^{-2} \), we observe that the mobilities are approximately \( \mu^{\text{double}} \approx 8252 \ (\text{cm}^2/\text{Vs}) \), and \( \mu^{\text{single}} \approx 9073 \ (\text{cm}^2/\text{Vs}) \). In conclusion, we can enhance the mobility of an infinite square quantum well by pairing two single quantum wells such that they are separated by \( d > 100 \ \text{\AA} \) and the widths of the two wells differ (either \( L_2 = L_1 + k \ (\text{\AA}) \) or \( L_1 = L_2 + k \ (\text{\AA}) \)) with \( k < 30 \ \text{\AA} \). More specifically with \( k = 10 \ \text{\AA} \), \( L_1 = 40 \ \text{\AA} \), \( L_2 = 30 \ \text{\AA} \), \( d = 100 \ \text{\AA} \), \( N_s = 10^{12} \ \text{cm}^{-2} \), \( T = 77 \ \text{K} \), \( \mu^{\text{double}} \approx 4.5 \times \mu^{\text{single}} \).

Figure \ref{fig6} illustrates the mobility trends of single and double quantum wells based on two distinct screening functions: \( \frac{1}{\varepsilon^2(q,T)} \) and \( \left( \frac{W_{ij}(q,T)}{V_{ij}(q)} \right)^2 \). In this analysis, the density of the single quantum well is adjusted to mirror that of the first-layer quantum well in the double-layer configuration, under conditions of \( T = 77 \ \text{K} \) and \( L = 140 \ \text{\AA} \). The separation \( d \) between the two quantum wells is examined at values of \( d = 10 \ \text{\AA} \) and \( 100 \ \text{\AA} \). Notably, in Figures~6a and 6b, \( N_1 \) is defined as \( N_2 \) multiplied by the scaling factor ``\( a \)'', \( N_1 = N_2 \times a \); while in Figs. \ref{fig6}(c) and (d), the relationship is reversed, with \( N_2 = N_1 \times a \).

The findings are striking: LO scattering remains unaffected by the screening function, leading to consistent mobility values. Throughout the examined density range, the mobility of the single quantum well clearly surpasses that of the double quantum well. However, as the distance \( d \) increases to \( 100 \ \text{\AA} \), an intriguing shift occurs; only the RI scattering reveals that the double quantum well achieves greater mobility than its single counterpart. Nevertheless, the overall mobility of the double quantum well still registers slightly lower than that of the single quantum well, particularly in scenarios where \( N_1 = N_2 \times a \) (illustrated in Figure~6b). This analysis underscores the nuanced dynamics of quantum well mobility and emphasizes the resilience of single quantum wells under varying conditions.

\section{Conclusion}

The theoretical calculations presented here decisively identify the highest possible mobility of the AlGaN/GaN/AlGaN infinite square quantum well. This paper clearly outlines a robust theoretical approach to enhance mobility by effectively coupling two quantum wells and positioning them in air. The total mobility incorporates a range of scattering mechanisms, including static potential scattering, negative phonon elastic scattering, and inelastic scattering from longitudinally polarized optical phonons (LO phonons). We systematically compare the mobility of the double quantum well, calculated using the double-layer screening function $\left( \frac{W_{ij}(q,T)}{V_{ij}(q)} \right)^2$, against that of the single quantum well, using the single-layer screening function $\frac{1}{\varepsilon^2(q,T)}$. It is critical to emphasize that the mobility affected by LO scattering remains independent of the shielding function. At room temperature ($T = 300\,\mathrm{K}$), LO scattering primarily determines total mobility, confirming that the mobility of a single quantum well consistently exceeds that of a double quantum well. When we examine conditions in which the density and well width of both quantum wells increase equally, we find that at a distance $d = 10\,\text{\AA}$ between the two wells, the mobility of the single quantum well remains superior. However, it is important to recognize that as the distance $d$ increases, the total mobility of the double quantum well overtakes that of the single quantum well when $d > 40\,\text{\AA}$. In scenarios where the density and well width of the two quantum wells vary independently, $d = 100\,\text{\AA}$, the double quantum well's mobility decisively surpasses that of the single quantum well when $L_2 = L_1 + k\,(\text{\AA})$ or $L_1 = L_2 + k\,(\text{\AA})$ with $k < 30\,\text{\AA}$, ensuring that either $L_1$ or $L_2$ is below $30\,\text{\AA}$. These theoretical calculations build solidly upon previous work and are instrumental for enhancing the mobility of semiconductor materials, a crucial goal for manufacturers today. We are committed to advancing these theoretical calculations further, exploring the impact of different dielectrics between the two quantum wells, and investigating the potential of utilizing GaN in the second quantum well while employing a material with a zincblende crystal structure.

\section*{Conflict of Interest}
The authors declare that they have no conflict of interest.

\bibliographystyle{plain}
\bibliography{references}  

\end{document}